\documentstyle[12pt,fleqn]{article}
\begin{document}
\baselineskip=7.5mm

\begin{center}
{\large\bf $\gamma$-Ray Bursts and Afterglows from Rotating Strange Stars
     and Neutron Stars}

\vspace{5.0mm}
Z. G. Dai and T. Lu

{\em Department of Astronomy, Nanjing University, Nanjing 210093, China}

\end{center}

\vspace{5mm}

{\small
We here discuss a new model of $\gamma$-ray bursts (GRBs) 
based on differentially rotating strange stars. 
Strange stars in this model and differentially rotating neutron stars in the
Klu\'zniak-Ruderman model can produce extremely relativistic,
variable fireballs required by GRBs and then become millisecond pulsars.
The effect of such pulsars on expansion of the postburst fireballs through
magnetic dipole radiation is studied. We show that these two models can 
explain naturally not only various features of GRBs but also light curves
of afterglows.  }

\vspace{2mm}

\vspace{8mm}
$\gamma$-ray bursts (GRBs) has been one of the greatest mysteries 
in high-energy astrophysics for about thirty years.
The observations of GRBs and a few afterglows [1-3] 
show that their sources are at cosmological distances. 
These observations require that any energy-source model 
should satisfy the following general features: 
(i) The model should produce an extremely relativistic fireball containing 
an amount of energy $\approx 10^{51}$--$10^{52}$\,ergs [4].
(ii) The high Lorentz factor of the fireball implied by the short variability
and the non-thermal spectra [5] requires that the fraction of 
contaminating baryons be less than 1\% of the explosive energy
(viz., $\le 10^{-5}M_\odot$).  (iii) The fireball is 
highly variable (required by the multi-peak feature
of the light curves of GRBs) and it should last for the GRB duration
(typically a few dozen seconds), implying that the source cannot release 
energy only once [4,6].  (iv) The fireball should be
a rare event occurring about once per million years in a galaxy [4]. 
(v) Finally, the model should account for
the decline-rise-decline behavior of the R-band light curve of 
the afterglow from GRB 970508 [7,8].
In this Letter, we combine features of some existing pulsar GRB models,
build upon them, and synthesize a new model of GRBs based on differentially
rotating strange stars.

Recently Klu\'zniak and Ruderman [9] argued differentially
rotating neutron stars as an origin of GRBs:
In a differentially rotating neutron star, internal poloidal
magnetic field ($B$) will be wound up into a toroidal configuration
and linearly amplified as one part of the star rotates about the other part.
Only when it increases up to a critical field value, $B_f$,
will the toroidal field be sufficiently buoyant to overcome fully
the stratification in neutron star composition. And then the buoyant
magnetic torus will be able to float up to break through the stellar surface.
Reconnection of the newborn surface magnetic field will lead to a quickly
explosive event with a large amount of energy, which could be a sub-burst
of a GRB. This model accounts well for the typical fluence in
each observed sub-burst (peak), for the number of peaks, for the time
interval between peaks, and for the rapid rise times and variability.
This model may result in low-mass loading baryonic matter.

Now let's discuss a new model of GRBs based on rotating
strange stars in detail. Strange stars have been widely 
studied [10,11]. However, two objections against the existence 
of isolated strange stars (with mass $\sim 1.4M_\odot$)
result from astrophysical arguments. On the one hand, it has been argued
that the disruption of a single strange star can
contaminate the entire galaxy, and thus all {\em neutron} stars have
in fact converted to strange stars [12]. Furthermore,
if strange stars can be created directly in supernovae, then some
strange stars must have more massive compact companions, and merging of
such binaries will lead to the Madsen-Caldwell-Friedman effect [13].
This result conflicts with the postglitch behavior of pulsars which
has been described well by the neutron-superfluid vortex creep
theory, rather than by the strange-star models [14]. On the other hand,
an isolated neutron star with a stiff equation of state (EOS) and 
with mass of $\sim 1.4M_\odot$ cannot convert to a strange star, 
because its central density is lower than the deconfinement density 
of neutron matter $\sim\,$(6--9)$\rho_0$ (here $\rho_0$ is the nuclear-matter 
density) [15]. 

However, there are two scenarios by which strange stars are born,
and which don't conflict with the above two arguments. First, it has 
been proposed [16] that when neutron stars in low-mass x-ray binaries 
accrete sufficient mass, they may convert to strange stars. The conversion 
should proceed in a detonation mode (at the speed of sound) [17], 
and thus the timescale for this process  is about 0.1\,ms. The direct
criterion for the conversion is that the total
masses of accreting neutron stars exceed $\sim 1.8M_\odot$ if the EOS
at high density is rather stiff. Fortunately, the theoretical
analysis shows that the amount of matter accreted by the 
radio pulsars in many binary systems exceeds $\sim 0.5M_\odot$ [18]. 
If this is true, some of these pulsars should have masses over
$1.8M_\odot$. The observation of kHz quasi-periodic
oscillations in eight low-mass x-ray binaries and the analysis of
high-resolution optical spectroscopy of Cyg X-2 indicate that the masses
of neutron stars in these systems are larger than $1.8M_\odot$ [19].

We consider an alternative scenario. It is usually thought
that due to gravitational radiation a neutron-star binary will
eventually merge to a black hole. This results from the argument
that the EOS for neutron matter at high density is soft.
But, the EOSs in neutron stars are thought to be moderately stiff to
stiff for the following reasons: (i) Soft EOSs at high densities are ruled
out by the postglitch recovery in four pulsars [20]. (ii)
The high masses ($\ge 1.8M_\odot$) of the neutron stars in a few low-mass
x-ray binaries also rule out soft EOSs [19]. (iii) The soft EOS such as 
kaon condensation doesn't seem to occur in stable neutron stars [21].
If the EOS for neutron matter is sufficiently stiff, therefore, the
post-merger objects of Hulse-Taylor-like binaries may be massive neutron
stars rather than black holes.
The same outcome would be achieved if the initial masses of
the merging neutron stars were low, e.g. $M\sim 1M_\odot$. According
to the first scenario, these massive neutron stars will subsequently
convert to strange stars.

If strange stars form in some x-ray binaries by these scenarios,
we now show that such stars are very difficult to be disrupted.
The total number of strange stars in a galaxy is about $10^4$ and
their number in globular clusters (GCs) is $N_{ss}\sim 10^3$.
The neutron star number density in GCs is $n_{ns}\sim 3\cdot 10^{-59}
(R_g/10{\rm kpc})^{-3}\xi\,{\rm cm}^{-3}$, where $R_g$ is the galactic radius
and $\xi\sim 10$, and we have assumed that the total neutron-star number
is $\sim 10^8$. The cross-section for a neutron star capturing a strange
star is usually approximated by $\sigma \sim 4\pi G^2M^2/v^4$,
where $M$ is the mass of the neutron star and $v$ its velocity.
Therefore, we find the rate for neutron star-strange star capture events
in GCs, $\Re \sim \sigma vn_{ns}N_{ss}\sim 10^{-17}(M/M_\odot)^2(v/500
{\rm km}\,{\rm s}^{-1})^{-3} (R_g/10{\rm kpc})^{-3} 
(\xi/10)\,{\rm yr}^{-1}$, and the rate outside GCs should be smaller.
The actual rate for the strange stars to be first captured and
then disrupted should be much less than $\Re$ due to the requirement of
massive neutron stars whose number is rather small.

For the above two scenarios, the pre-conversion neutron stars all have
periods of the order of 1 ms because of accretion-induced spin-up in
low-mass x-ray binaries and synchrotronization during the coalescence
of neutron-star binaries. Since the moments of inertia always decrease
during the conversion, the resulting strange stars will rotate
at periods $\le 1\,$ms. Furthermore, differential rotation may
occur in the interiors of these newborn strange stars due to the fact that
the density profile of a strange star is much different from that of
a neutron star with the same mass [11]. According to the basic idea of
the Klu\'zniak-Ruderman model, such differentially rotating strange stars
could lead to a series of sub-bursts of GRBs.

Our model can account for the following features. First, {\em
sufficient energy}. The total available energy released in our model should
be the differential rotation energy, viz., $E=\hat{I}\Omega^2/2
=5\cdot 10^{51}\,{\rm ergs}\,\hat{I}_{44}\Omega_4^2$,
where $\hat{I}$ is the effective moment of inertia of differential
rotation ($\hat{I}_{44}=\hat{I}/10^{44}{\rm g}\,{\rm cm}^2$), and
$\Omega$ the effective difference in angular velocity of rotation
($\Omega_4=\Omega/10^4{\rm s}^{-1}$). We here estimate the critical field
based on $B_f^2/(8\pi)=f\rho c_s^2$, where $f$ is the
dimensionless parameter accounting for the stratification ($\sim 1\%$ for
massive strange stars), $\rho$ the typical mass density of the strange star, 
and $c_s$ the speed of sound ($c_s=c/\sqrt{3}$ for strange matter). 
Thus, we obtain $B_f\approx 2\cdot 10^{17}(\rho/4\cdot 10^{14}{\rm g}
\,{\rm cm}^{-3})^{1/2}\,{\rm G}$.

Second, {\em very-low-mass baryon contamination}. Because the strange
star has a thin crust ($\sim 10^{-5}M_\odot$) [10,11],
the resulting fireball in our model is almost free of baryons,
alleviating the baryon-contamination problem. This point was raised 
by Haensel et al. [22], who considered merging of strange stars
as an origin of GRBs. Of course, this problem can also be avoided 
in many models [23-25], where power is extracted from rapidly 
spinning neutron stars or black holes with strong magnetic fields by 
Poynting flux, which are later converted to $\gamma$-rays and 
lower-frequency photons. But, it is unclear whether these models can 
also explain the spiky light curves of GRBs.

Third, {\em temporal property}. Following [9],
we estimate the time interval between sub-bursts,
$\tau_i=(2\pi/\Omega)(B_f/B)\approx 10\,{\rm s}B_{13}^{-1}\Omega_4^{-1}$,
and the total duration, $\tau_d\approx 60\,{\rm s}B_{13}^{-1}\Omega_4$,
where $B_{13}$ is in units of $10^{13}\,$G. The internal field
could be generated by either field-freezing or by a dynamo.
Duncan and Thompson [23] have shown that whether a dynamo operates due
to convection driven by a large neutrino flux is dependent on
the Rossby number $N_R=P/\tau_{conv}$, where $P$ is the stellar period 
and $\tau_{conv}$ the convection overturn timescale at the base
of the convection zone. If $N_R\le 1$, an efficient dynamo can
result and the (poloidal) field grows exponentially to $B\sim 10^{15}\,$G;
if $N_R>1$, a dynamo is suppressed. It means that if a dynamo has 
operated, the duration is $\tau_d\approx 0.6\,{\rm s}
(B_{13}/10^2)^{-1}\Omega_4$, leading to short bursts; but if not, 
long bursts will form. This provides an explanation for the bimodal 
duration distribution of GRBs. This distribution has also been 
proposed to be associated with submillisecond neutron stars through
gravitational instability [25].

Fourth, {\em low burst rate}. It has been argued [16,26]
that the rate of conversion of accreting neutron stars in low-mass
x-ray binaries, $\sim 10^{-6}$/yr per galaxy, is consistent with
the observed GRB rate. In addition, the rate of coalesence of
Hulse-Taylor-like binaries has been shown to correspond closely to
the observed GRB rate [27].

Both the Klu\'zniak-Ruderman model and our model can 
produce an extremely relativistic, variable fireball, which will 
lead to internal shocks [28] and subsequent external shocks [29]. 
A GRB will be produced once the kinetic energy is dissipated and radiated as
$\gamma$-rays through synchrotron or possibly inverse-Compton emission from
the accelerated electrons in these shocks. At the same time, the strange
star is becoming a pulsar. Therefore, two common by-products of these models 
are a strongly magnetic millisecond pulsar and a postburst relativistic 
fireball. It is natural to expect that the central pulsar affects 
the evolution of the postburst fireball and the afterglow from this 
fireball. Can this effect explain the observed afterglows? Yes. 
Here we would provide a case in which this effect can explain well 
the optical afterglow of GRB 970508.

At the center of the fireball, the pulsar loses its rotational energy 
through magnetic dipole radiation, whose power varies
with time as $L(t) \propto (1+t/T)^{-2}$,
where $t$ is one measure of time in the burster's rest frame,
and $T$ the initial spin-down timescale. For $t< T$, $L$ can be
thought as a constant; but for $t\gg T$, $L$ decays as $\propto t^{-2}$.

The pulsar radiates electromagnetic waves with frequency of $\omega=2\pi/P
\sim 10^4\,{\rm s}^{-1}$. These waves are absorbed by the shocked 
interstellar medium (ISM) because the plasma frequency of 
the shocked ISM is much higher than $\omega$. This implies 
that the pulsar pumps continuously its rotational energy into 
the shocked ISM. We assume that the expansion of the fireball in
uniform ISM is relativistic and adiabatic. At a time $t$, the shocked
ISM energy is given by $4\pi r^2(r/4\gamma^2)\gamma^2e'$, 
where $r\approx ct$ is the blastwave radius, $\gamma$
the Lorentz factor of the fireball, and $e'=4\gamma^2nm_pc^2$ the shocked
ISM energy density in the comoving frame (where $n$ is the electron number 
density of the unshocked ISM) [30]. This energy should be equal
to the sum of half of the initial energy ($E/2$) and the energy which
the fireball has obtained from the pulsar based on energy conservation:
\begin{equation}
4\pi nm_pc^2\gamma^2r^3=\frac{E}{2}+\int_0^t (1-\beta)L(t-r/c)dt\,,
\end{equation}
where $\beta=(1-1/\gamma^2)^{1/2}$. Because the Lorentz factor of
the fireball at the initial stage decays as $\gamma\propto
t^{-3/2}$, the timescale in which the fireball has obtained energy $\sim E/2$
from the pulsar is estimated as $4\gamma^2E/L$, measured in
the burster's rest frame. The corresponding observer-frame timescale ($\tau$)
is equal to this timescale divided by $2\gamma^2$, viz., 
$\tau=2E/L$. We assume $\tau\ll T$, or $P <
1.6\,{\rm ms}(E_{51}/4)^{-1/2}I_{45}^{1/2}$. Clearly, this
inequality is independent on the stellar magnetic field.
Please note that Blackman and Yi [31] recently discussed
the dissipation of the rotational energy of a pulsar by large amplitude
electromagnetic waves, which are directly converted to an afterglow, 
but in our model, afterglows result from shocks absorbing such waves.

We consider below the synchrotron radiation from the accelerated electrons 
behind the shock, and assume a power-law distribution of the 
electrons with index $p$. We analyze evolution of the afterglow:
(i) At the initial stage, viz., observer-frame time $t_\oplus\ll \tau$,
the second term on the right side of equation (1) can be neglected and 
the Lorentz factor of the fireball decays as
$\gamma\propto t_\oplus^{-3/8}$. In this case, we have found that
the fireball will rapidly go into the slow cooling phase, and
the radiation at optical band will come from those radiative 
electrons a few hours later after the initial burst.
This conclusion is similar to that of the recent studies [32,33].
Therefore, the optical flux from the fireball
$F_\nu\propto \nu^{-p/2}t_\oplus^{(2-3p)/4}$.
The result $p\approx 1.0$ inferred from the observed spectrum index
in two days following the initial burst [8] leads to
the R-band flux $\propto t_\oplus^{-1/4}$. This satisfactorily
accounts for the slow decline of the R-band light curve in $\sim 8\,$ 
hours after the burst [8]. 

(ii) For $T>t_\oplus\gg \tau$, the term $E/2$ in equation (1) 
can be neglected and the Lorentz factor of the fireball decays
as $\gamma\propto t_\oplus^{-1/4}$. In this case, 
the comoving-frame equipartition magnetic field decreases as
$B'\propto \gamma\propto t_\oplus^{-1/4}$, and the synchrotron
break frequency drops in time as $\nu_m\propto \gamma^3B'\propto
t_\oplus^{-1}$. At the same time, since the comoving electron
number density is $n'_e\propto \gamma\propto t_\oplus^{-1/4}$ and
the comoving width of the emission region $\Delta\! r' \sim r/\gamma
\propto t_\oplus^{3/4}$, according to M\'esz\'aros \& Rees in [7],
the comoving intensity $I'_\nu\propto
n'_eB'\Delta\! r'\propto t_\oplus^{1/4}$. So the observed peak flux density
increases in time based on $F_{\nu_m}\propto t_\oplus^2 \gamma^5I'_{\nu_m}
\propto t_\oplus$. For a power-law distribution, electrons with different
Lorentz factors may have different radiative efficiencies.
Sari et al. [32] have given a critical value, $\gamma_c\propto 
\gamma^{-3}t_\oplus^{-1}\propto t_\oplus^{-1/4}$, 
above which cooling by synchrotron radiation is significant.
The corresponding synchrotron frequency decays as $\nu_c\propto
\gamma B'\gamma_c^2\propto t_\oplus^{-1}$. From [32], we find
the optical flux 
$F_\nu= (\nu_c/\nu_m)^{-(p-1)/2} (\nu/\nu_c)^{-p/2}F_{\nu_m}
      \propto \nu^{-p/2}t_\oplus^{(2-p)/2}$,
and therefore, $p\approx 1.0$ further leads to the result that the optical
flux increases with time, being in agreement with the subsequent rise of 
the R-band light curve [8]. 

(iii) For $t_\oplus\gg T$, the power of the pulsar due to magnetic 
dipole radiation rapidly decreases as $L\propto t^{-2}$, and
the evolution of the fireball is hardly influenced by the stellar radiation. 
As in stage (i), the optical flux is $F_\nu\propto \nu^{-p/2}t_\oplus
^{(2-3p)/4}$. Because $p\approx 2.2$, inferred from the spectrum
index observed two days later after the burst, the R-band flux $\propto 
t_\oplus^{-1.2}$, which is also well consistent with the observations [8].

The effect of millisecond pulsars on afterglows can also be shown
schematically in Figure 1. We further give constraints on stellar
parameters. According to the definitions of $\tau$ and $T$,
we obtain the stellar period,
$P=0.6\,{\rm ms}(E_{51}/4)^{-1/2}(T/2{\rm d})^{-1/2}
           (\tau/0.4{\rm d})^{1/2}I_{45}^{1/2}$,
and the magnetic field strength
$B_{13}=3.1(E_{51}/4)^{-1/2}(T/2{\rm d})^{-1}(\tau/0.4{\rm d})^{1/2}
       I_{45}R_6^{-3}$,
where $E_{51}=E/10^{51}{\rm ergs}$, $I_{45}$ is the moment of inertia
in $10^{45}{\rm g}\,{\rm cm}^2$ and $R_6$ the stellar radius
in $10^6\,$cm.
The observations indicate $\tau\sim 0.4\,$days and $T\sim 2\,$days
[8]. If taking $(E_{51}/4)^{-0.5}I_{45}R_6^{-3}
\sim 1$, we have $B_{13}\sim 3.1$, and the total duration of the main burst
($t_d$) is estimated to be about 20 seconds. This is in excellent agreement 
with the observation by the BeppoSAX satellite [34]. In addition, we find
$P\approx 0.6\,$ms, implying that the pulsar has a submillisecond period. 
What we want to point out is that such a period for the formation scenario 
of a strange star in a low-mass x-ray binary may be due to preconversion 
accretion-induced spin-up and due to decrease of stellar moment 
of inertia during conversion, and a magnetic field $\sim 10^{13}$\,G may 
result from preconversion crustal plate motion and/or from a dynamo in
the strange star.

If a GRB results from magnetic reconnection on the surface of a
differentially rotating strange star or neutron star, we cannot 
clearly determine the region in which such a process takes place. 
A beaming of the emission might form if this process occurs 
on a small area of the stellar surface as solar flares. This might 
provide a solution to the large-energy problem of GRB 971214.

Finally, it is interesting to note that two other kinds of high-energy 
transients, the soft $\gamma$-ray repeaters and the hard x-ray burster 
(GRO J1744-28), have also been described well by the strange-star models 
in which the stellar crust-breaking leads to bursts [35].

We thank the referees for their enlightening comments and valuable 
suggestions. This work was supported by the National Natural Science 
Foundation of China.

\vspace{4mm}

\baselineskip=5mm

\begin{center}
{\bf References}
\end{center}

\begin{description}
\item $[1]$ G. J. Fishman and C. A. Meegan,
Ann. Rev. Astron. Astrophys. {\bf 33}, 415 (1995).
\item $[2]$ J. van Paradijs {\em et al.}, Nature (London) {\bf 386}, 686 (1997).
\item $[3]$ M. R. Metzger {\em et al.}, Nature (London) {\bf 387}, 878 (1997).
\item $[4]$ For a recent review see T. Piran, astro-ph/9801001.
\item $[5]$ E. Woods and A. Loeb, Astrophys. J. {\bf 453}, 583 (1995).
\item $[6]$ R. Sari and T. Piran, Astrophys. J. {\bf 485}, 270 (1997).
\item $[7]$ Simple analytical models are successful at explaining the 
            overall power-law decay behavior of light curves 
            [e.g., P. M\'esz\'aros and M. J. Rees, Astrophys. J. {\bf 476}, 
            232 (1997); E. Waxman, {\em ibid.} {\bf 485}, L5 (1997); 
            M. Vietri, {\em ibid.} {\bf 488}, L105 (1997);
            R. Wijers, M. J. Rees and P. M\'esz\'aros, Mon. Not. R. Astron.
            Soc. {\bf 288}, L51 (1997)]. But, the optical and x-ray light
            curves of the observed afterglows have been shown to depart from 
            this behavior [8].
\item $[8]$ S. G. Djorgovski {\em et al.}, Nature (London) {\bf 387}, 
             876 (1997); A. J. Castro-Tirado, {\em et al.}, Science 
             {\bf 279}, 1011 (1998); H. Pedersen {\em et al.}, 
             Astrophys. J. {\bf 496}, 311 (1998);
             T. J. Galama {\em et al.}, {\em ibid.} {\bf 497}, L13 (1998).
\item $[9]$ W. Klu\'zniak and M. Ruderman, Astrophys. J. {\bf 505}, 
              L113 (1998).
\item $[10]$ C. Alcock, E. Farhi and 
             A. Olinto, Astrophys. J. {\bf 310}, 261 (1986); 
             P. Haensel, J. L. Zdunik and R. Schaeffer, 
             Astron. Astrophys. {\bf 160}, 121 (1986); N. K. Glendenning 
             and F. Weber, Astrophys. J. {\bf 400}, 647 (1992); Y. F. Huang 
             and T. Lu, Astron. Astrophys. {\bf 325}, 189 (1997).
\item $[11]$ N. K. Glendenning, {\em Compact Stars: Nuclear Physics, Particle
              Physics and General Relativity}, (Springer-Verlag,
              New York, 1997), pp.337-361.
\item $[12]$ J. Madsen, Phys. Rev. Lett. {\bf 61}, 2909 (1988);
             R. R. Caldwell and J. L. Friedman, Phys. Lett. B {\bf 264},
             143 (1991).
\item $[13]$ W. Klu\'zniak, Astron. Astrophys. {\bf 286}, L17 (1994).
\item $[14]$ M. A. Alpar, Phys. Rev. Lett. {\bf 58}, 2152 (1987).
\item $[15]$ G. Baym, in {\em Neutron Stars: Theory and observation}, 
              J. Ventura and D. Pines, eds. (Kluwer, Dordrecht, 1991), p.21.
\item $[16]$ K. S. Cheng and Z. G. Dai, Phys. Rev. Lett. {\bf 77}, 
              1210 (1996).
\item $[17]$ G. Lugones, O. G. Benvenuto and H. Vucetich, Phys. Rev. D
               {\bf 50}, 6100 (1994).
\item $[18]$ E. P. J. van den Heuvel and O. Bizaraki, Astron. Astrophys. 
              {\bf 297}, L41 (1995).
\item $[19]$ W. Zhang, T. E. Strohmayer and J. H. Swank, Astrophys. J. 
              {\bf 482}, L167 (1997); P. Kaaret, E. C. Ford and K. Chen, 
              {\em ibid.} {\bf 480}, L27 (1997);
              J. Casares, P. Charles and E. Kuulkers,  
              {\em ibid.} {\bf 493}, L39 (1998).
\item $[20]$ B. Link, R. I. Epstein and K. A. Van Riper, Nature (London) 
              {\bf 359}, 616 (1992). 
\item $[21]$ V. R. Pandharipande, C. J. Pethick and V. Thorsson, 
              Phys. Rev. Lett. {\bf 75}, 4567 (1995);
              Z. G. Dai and K. S. Cheng, Phys. Lett. B {\bf 401}, 
              219 (1997).
\item $[22]$ P. Haensel, B. Paczy\'nski and P. Amsterdamski, Astrophys. J.
             {\bf 375}, 209 (1991).
\item $[23]$ R. C. Duncan and C. Thompson, Astrophys. J. {\bf 392}, L9 
             (1992). 
\item $[24]$ V. V. Usov, Nature (London) {\bf 357}, 452 (1992);
             C. Thompson, Mon. Not. R. Astron. Soc. {\bf 270}, 480 (1994);
             E. G. Blackman, I. Yi and G. B. Field, Astrophys. J. {\bf 473},
             L79 (1996); P. M\'esz\'aros and M. J. Rees, {\em ibid.} 
             {\bf 482}, L29 (1997).
\item $[25]$ I. Yi and E. G. Blackman, Astrophys. J. {\bf 494}, L163 (1998). 
\item $[26]$ K. S. Cheng and Z. G. Dai, Astrophys. J. {\bf 492}, 281 (1998).
\item $[27]$ B. Paczy\'nski, Acta Astron. {\bf 41}, 257 (1991); 
             R. Narayan, T. Piran and A. Shemi, Astrophys. J. 
             {\bf 379}, L17 (1991); E. S. Phinney, {\em ibid.} {\bf 380},
             L17 (1991).
\item $[28]$ M. J. Rees and P. M\'esz\'aros, Astrophys. J. {\bf 430}, 
             L93 (1994); B. Paczy\'nski and G. Xu, {\em ibid.} {\bf 427}, 
             708 (1994).
\item $[29]$ M. J. Rees and P. M\'esz\'aros, Mon. Not. R. Astron. Soc.
             {\bf 258}, 41p (1992).
\item $[30]$ R. D. Blandford and C. F. McKee, Phys. Fluids {\bf 19}, 
              1130 (1976).
\item $[31]$ E. G. Blackman and I. Yi, Astrophys. J. {\bf 498}, L31 (1998).
\item $[32]$ R. Sari, T. Piran and R. Narayan, Astrophys. J. {\bf 497}, L17
             (1998).
\item $[33]$ A. Panaitescu, P. M\'esz\'aros and M. J. Rees, astro-ph/9801258.
\item $[34]$ L. Piro {\em et al.}, Astron. Astrophys. {\bf 331}, L41 (1998).
\item $[35]$ K. S. Cheng and Z. G. Dai, Phys. Rev. Lett. {\bf 80}, 18 
             (1998); K. S. Cheng, Z. G. Dai, D. M. Wei and T. Lu,
              Science {\bf 280}, 407 (1998).
\end{description}

\baselineskip=7.5mm
\vspace{10mm}

\begin{center}
{\bf FIGURE CAPTION}
\end{center}

\vspace{2mm}

\noindent
FIG. 1. Afterglow light curves with a pulsar (solid line) and
without a pulsar (dashed line).

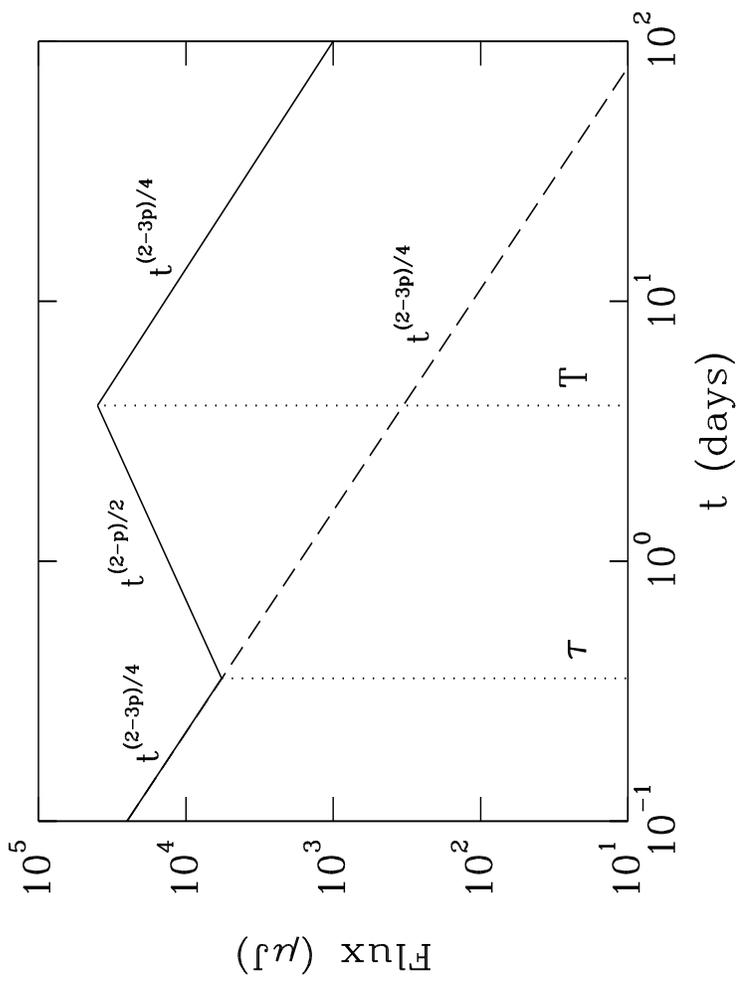
\end{document}